\documentclass[notitlepage,superscriptaddress,twocolumn,prx,nofootinbib]{revtex4-2}

\expandafter\let\csname equation*\endcsname\relax
\expandafter\let\csname endequation*\endcsname\relax
\expandafter\let\csname eqnarray*\endcsname\relax
\expandafter\let\csname endeqnarray*\endcsname\relax
\usepackage[ngerman,english]{babel}
\addto\captionsenglish{}
\usepackage[utf8]{inputenc}
\usepackage{array}
\usepackage{amsmath}
\usepackage{amssymb}
\usepackage{graphicx}
\usepackage{braket}
\usepackage{color}
\usepackage{float}
\usepackage{soul}
\usepackage{enumitem}
\usepackage[dvipsnames]{xcolor}
\usepackage{comment}
\usepackage{bbold}

\usepackage{hyperref}

\newcommand{\ie}{i.\,e.\ }
\newcommand{\cf}{cf.\ }
\newcommand{\eg}{e.\,g.\ }
\newcommand{\dd}{\mathrm{d}}
\newcommand{\re}{\operatorname{Re}}

\newcommand{\abs}[1]{\lvert#1\rvert}
\newcommand{\be}{\mathbf{e}}
\newcommand{\br}{\mathbf{r}}
\newcommand{\bk}{\mathbf{k}}

\newcommand{\bd}{\mathbf{d}}
\newcommand{\id}{\mathbb{1}}

\newcommand{\corr}{\mathrel{\scalebox{1}[.8]{$\widehat{=}$}}} 

\renewcommand{\rm}{\mathrm}

\begin{document}

\title{Observation of Light-Induced Dipole-Dipole Forces in Ultracold Atomic Gases}
\author{Mira Maiwöger}
\affiliation{Vienna Center for Quantum Science and Technology, Atominstitut, TU Wien, 1020 Vienna, Austria}
\author{Matthias Sonnleitner}
\affiliation{Institute for Theoretical Physics, University of Innsbruck, 6020 Innsbruck, Austria}
\affiliation{Erwin Schr\"{o}dinger International Institute for Mathematics and Physics, University of Vienna, 1090 Vienna, Austria}
\author{Tiantian Zhang}
\affiliation{Vienna Center for Quantum Science and Technology, Atominstitut, TU Wien, 1020 Vienna, Austria}
\author{Igor Mazets}
\affiliation{Vienna Center for Quantum Science and Technology, Atominstitut, TU Wien, 1020 Vienna, Austria}
\affiliation{Research Platform MMM ``Mathematics–Magnetism–Materials" \\  
c/o Fakult\"at f\"ur Mathematik, University of Vienna, 1090 Vienna, Austria}
\affiliation{Wolfgang Pauli Institute c/o Fakult\"at f\"ur Mathematik, University of Vienna, 1090 Vienna, Austria}
\author{Marion Mallweger}
\affiliation{Vienna Center for Quantum Science and Technology, Atominstitut, TU Wien, 1020 Vienna, Austria}
\affiliation{Department of Physics, Stockholm University, SE-106 91 Stockholm, Sweden}
\author{Dennis Rätzel}
\affiliation{Institut für Physik, Humboldt-Universität zu Berlin, 12489 Berlin, Germany}
\affiliation{Erwin Schr\"{o}dinger International Institute for Mathematics and Physics, University of Vienna, 1090 Vienna, Austria}
\author{Filippo Borselli }
\affiliation{Vienna Center for Quantum Science and Technology, Atominstitut, TU Wien, 1020 Vienna, Austria}
\author{Sebastian Erne}
\affiliation{Vienna Center for Quantum Science and Technology, Atominstitut, TU Wien, 1020 Vienna, Austria}
\author{Jörg Schmiedmayer}
\affiliation{Vienna Center for Quantum Science and Technology, Atominstitut, TU Wien, 1020 Vienna, Austria}
\author{Philipp Haslinger}
\email{philipp.haslinger@tuwien.ac.at}
\affiliation{Vienna Center for Quantum Science and Technology, Atominstitut, TU Wien, 1020 Vienna, Austria}
\affiliation{Erwin Schr\"{o}dinger International Institute for Mathematics and Physics, University of Vienna, 1090 Vienna, Austria}

\begin{abstract}
We investigate an attractive force caused by light induced dipole-dipole interactions in freely expanding ultracold $\mathrm{^{87}Rb}$ atoms. This collective, light-triggered effect results in a self-confining potential with interesting features: it exhibits nonlocal properties, is attractive for both red and blue-detuned light fields and induces a remarkably strong force that depends on the gradient of the atomic density. The experimental data are discussed in the framework of a theoretical model based on a local-field approach for the light scattered by the atomic cloud.
\end{abstract}

\maketitle
\section{Introduction}
A single atom interacting with a laser beam will usually experience a combination of radiation pressure pushing the atom along the beam as well as a dipole- or gradient force pulling the particle towards regions of a local beam intensity extremum~\cite{cohen1998nobel,phillips1998nobel}. In dense~\cite{zhang1994quantum,morice1995refractive,ruostekoski1997quantum,ruostekoski1997lorentz,krutitsky1999local} or periodically structured~\cite{shahmoon2017cooperative,shahmoon2019collective} ensembles, the collective back-action of the atoms on the light field can lead to significant additional effects~\cite{guerin2017light}. These include, for example, superradiance~\cite{inouye1999superradiant,uys2008cooperative}, modifications of emission patterns~\cite{jennewein2016coherent,jenkins2016collective}, or shifts of atomic resonance lines~\cite{javanainen2014shifts,pellegrino2014observation,hotter2019superradiant,glicenstein2020collective}. A number of theoretical proposals have discussed how this collective interaction and the resulting forces can reshape atomic clouds~\cite{zhang1994quantum,krutitsky1999local, mazets2000ground,giovanazzi2001one,giovanazzi2002density,odell2000bose}.

Here we report on the first observation of mechanical effects due to collective light-induced dipole-dipole (LI-DD) interactions without enhancement or selecting spatial mode structures by employing cavities~\cite{ritsch2013cold,griesser2013light,ostermann2016spontaneous,dimitrova2017observation}. We show that a homogeneously illuminated cloud of ultracold atoms experiences a remarkably strong compressing potential for both red and blue-detuned light fields. The resulting LI-DD potential minimum is intrinsically tied to the atomic ensemble and can freely evolve in additional external potentials.
These properties distinguish the LI-DD interaction from well known dipole forces or a previously reported effect termed electrostriction~\cite{matzliah2017observation}.

In simplified terms, the LI-DD interaction can be seen as a second-order effect where atoms interact with the light scattered by other particles. This way it effectively constitutes a nonlocal, self-confining, and controllable long-range particle-particle interaction~\cite{odell2000bose,giovanazzi2002density}.
The nonlocality of the LI-DD interaction manifests itself in the fact that the force does not depend only on the local atomic density. It could thus be a new way to tailor atomic interactions beyond $s$-wave scattering~\cite{inouye1998observation, vuletic1999observation,derrico2007feshbach,borselli2021two} or (static) dipole-dipole interaction in polar gases~\cite{lahaye2009physics,bottcher2020new}.
Our observations represent a first step towards an experimental implementation of various theoretical ideas based on the LI-DD potential properties~\cite{kurizki2004bose}. 

This work is organised as follows: In section~\ref{sec:S1-Experiment basics} we introduce the experimental setup, the measurement procedure, and give some theoretical intuition for the expected LI-DD effects. In section~\ref{sec:_Results} we present and discuss the experimental results. Section~\ref{sec:_Theory} gives more details on the theory and the numerical simulation. The work is closed by a short summary and discussion in section~\ref{sec:_Conclusion}. Further details on the experiment and numerical modelling is presented in the appendix.

\section{\label{sec:S1-Experiment basics}Experimental system}
\begin{figure*}
  \centering
      \includegraphics[width=\textwidth]{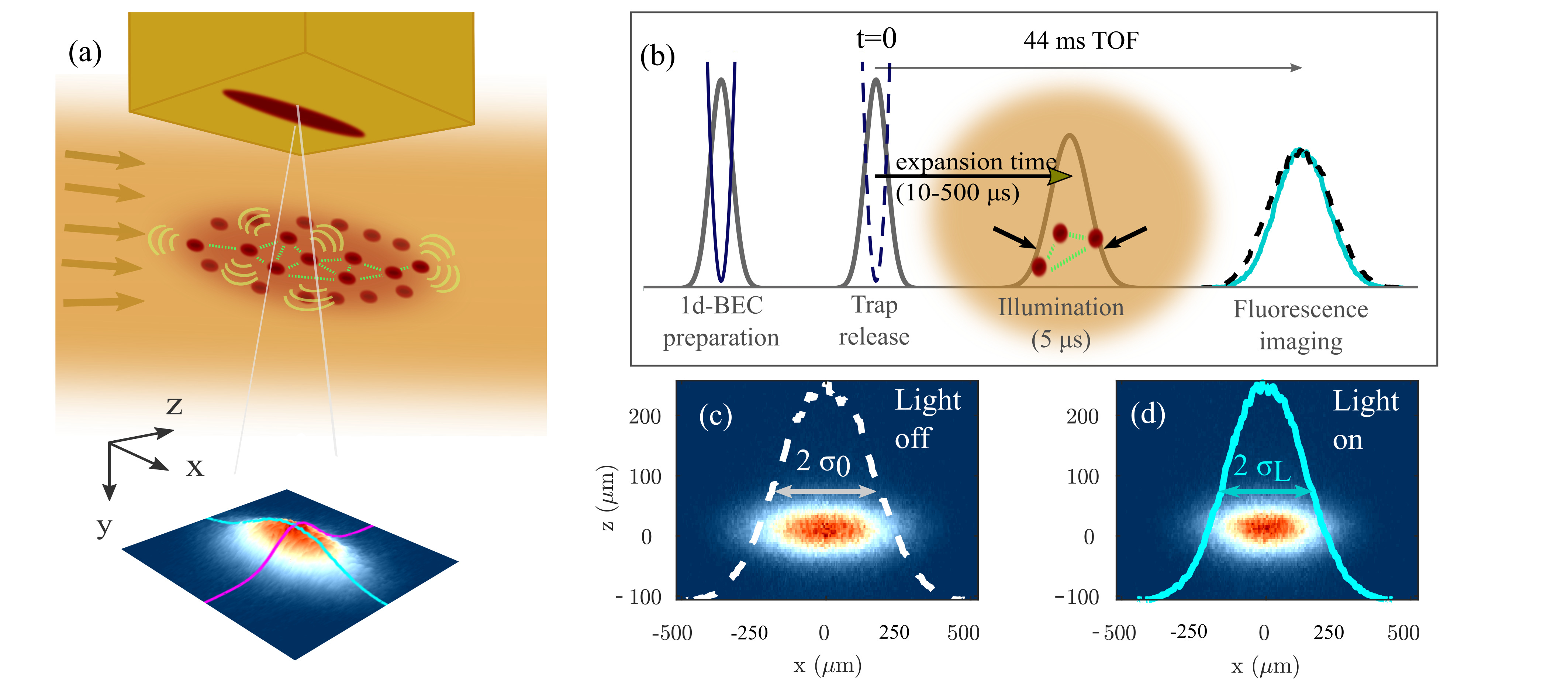}
    \caption{(a) 3D Illustration of the experimental setup. A 1d-BEC is magnetically trapped below an atom chip. After releasing the atoms from the trap, they are illuminated with a spatially homogeneous laser pulse to induce the LI-DD interaction. The beam is aligned nearly parallel to the long ($z$) axis (see text). After $44 \, \mathrm{ms}$ ToF the atomic cloud is imaged using a light sheet imaging system. (b) Sketch of experimental sequence: After switching off the trap, the 1d-BEC expands for $10$ to $500 \, \mathrm{\mu s}$ before being illuminated with a $5\,\mathrm {\mu s}$ long laser pulse. The LI-DD interaction causes the atomic cloud to contract in the transverse directions resulting in a reduced transverse width $\sigma_{L}$ compared to the width $\sigma_{0}$ without additional illumination of the freely expanding cloud. Averaged light sheet images with mean atom number $N = 6600(130)$ (c) without and (d) with illumination by the laser beam with blue detuned light ($\Delta = 100 \Gamma$) and intensity $I= 28.3(0.8)\, I_{\mathrm{sat}}$ for $5\,\mathrm{\mu s}$, $105\,\mathrm{\mu s}$ after trap release. The lines indicate the corresponding transverse density profile after integrating over the full extension of the 1d-BEC along the long $z$-axis. Note that the 1d-BEC expands mainly in the initially tightly confined transverse (radial) directions, resulting in an inverted aspect ratio of the cloud after $44\,\mathrm{ms}$ of flight in the light sheet image.}
    \label{fig:SETUP}
\end{figure*}
Our experimental setup (see Fig.~\ref{fig:SETUP} and Ref.~\cite{trinker2008multilayer,bucker2009single} for more details) is optimized to produce elongated one-dimensional $\mathrm{^{87}Rb}$ quasi-condensates (1d-BECs) of typically $N=(5 - 10 )\times 10^3$~atoms in the $F = 1$, $m_F=-1$ state, magnetically confined $60\,\mathrm{\mu m}$ below a gold coated atom chip \cite{trinker2008multilayer,folman2002microscopic}. The 1d-BEC is prepared in the radial ground state of a magnetic trap with axial frequency $\omega_z=2\pi \times 18.6\,\mathrm{Hz}$ and transverse frequency $\omega_{\perp}=2\pi\times 2.96\,\mathrm{kHz}$. In the trap, the atom density has an average Gaussian width of $a_\perp = \sqrt{\hbar/(m \omega_\perp)}\approx 200\,\mathrm{nm}$, with $m$ the atomic mass of $\mathrm{^{87}Rb}$. The length $L$ of the 1d-BEC in the axial direction depends on the number of atoms, for $N=7000$  atoms, $L \approx  90~\mathrm{\mu m}$ \cite{gerbier2004quasi}.
After switching off the magnetic trapping fields (in about $1\,\mathrm{\mu s}$) the atoms fall for $44\,\mathrm{ms}$ time-of-flight (ToF) until they reach the detection region (Fig.~\ref{fig:SETUP}(d)). There the atoms are imaged while passing through an on-resonant light sheet \cite{bucker2009single}.

$10 - 500\, \rm{\mu s}$ after switching off the magnetic trap, a $5\,\mathrm{\mu s}$ long laser pulse at various detunings and intensities is used to trigger the LI-DD interaction between the atoms. As the cloud expands rapidly in the transverse direction, changing the time delay between trap release and the laser pulse allows us to illuminate the sample at different mean atomic densities while keeping the total atom number constant. We choose a laser beam waist radius of $\sim 1\,\mathrm{mm}$, much larger than the size of the expanding 1d-BEC, to ensure homogeneous illumination and prevent residual dipole forces. The full experimental sequence is illustrated in Fig.~\ref{fig:SETUP}(b).
The laser beam inducing the LI-DD interaction propagates nearly parallel to the axial direction of the 1d-BEC. To avoid reflections on the vacuum window and the gold coated surface of the atom chip back onto the atom cloud, the beam is aligned with a horizontal angle of $\sim2^{\circ}$ and tilted downwards by $0.5^{\circ}$.

We probe the LI-DD interaction for mean atomic densities ranging from $260 - 3 \,\mathrm{atoms/\mu m^3}$, corresponding to free expansion times of the atomic cloud between $10 - 500\,\mathrm{\mu s}$ (see Fig.~\ref{fig:_ExpansionAfterCloud_ExpAndTheory}). The gravitational displacement of the cloud during the expansion amounts to less than $1\,\mathrm{\mu m}$ while its radius expands to less than $2\,\mathrm{\mu m}$ and remains much smaller than the waist of the laser beam.

\subsection{Theoretical intuition}\label{sec:_Theoretical_Intuition}
The concept of LI-DD interactions can be understood from two equivalent models: In a particle picture we first note that the well known radiation pressure and dipole forces arise from interaction between a single atom and an external light field. But particles immersed in a sufficiently large or dense cloud will also interact with light scattered by other atoms. This collective multiple scattering can be understood as an effective long-range particle-particle interaction.

Equivalently one can observe that atoms collectively reshape the incoming laser field through their refractive index and the resulting (local) electric field intensity gives rise to an effective atom-light potential~\cite{morice1995refractive,ruostekoski1997quantum,ruostekoski1997lorentz,krutitsky1999local}. This potential depends on the local as well as the integrated atomic density, see Eq.~\eqref{eq:_Val}. 

A gradient in the density thus leads to a gradient in the potential and results in a force on the atoms. As we will discuss in more detail in section~\ref{sec:_Theory}, this force turns out to be net attractive for both red and blue-detuned light fields.

\begin{figure*}
    \centering
    \includegraphics[width=\textwidth]{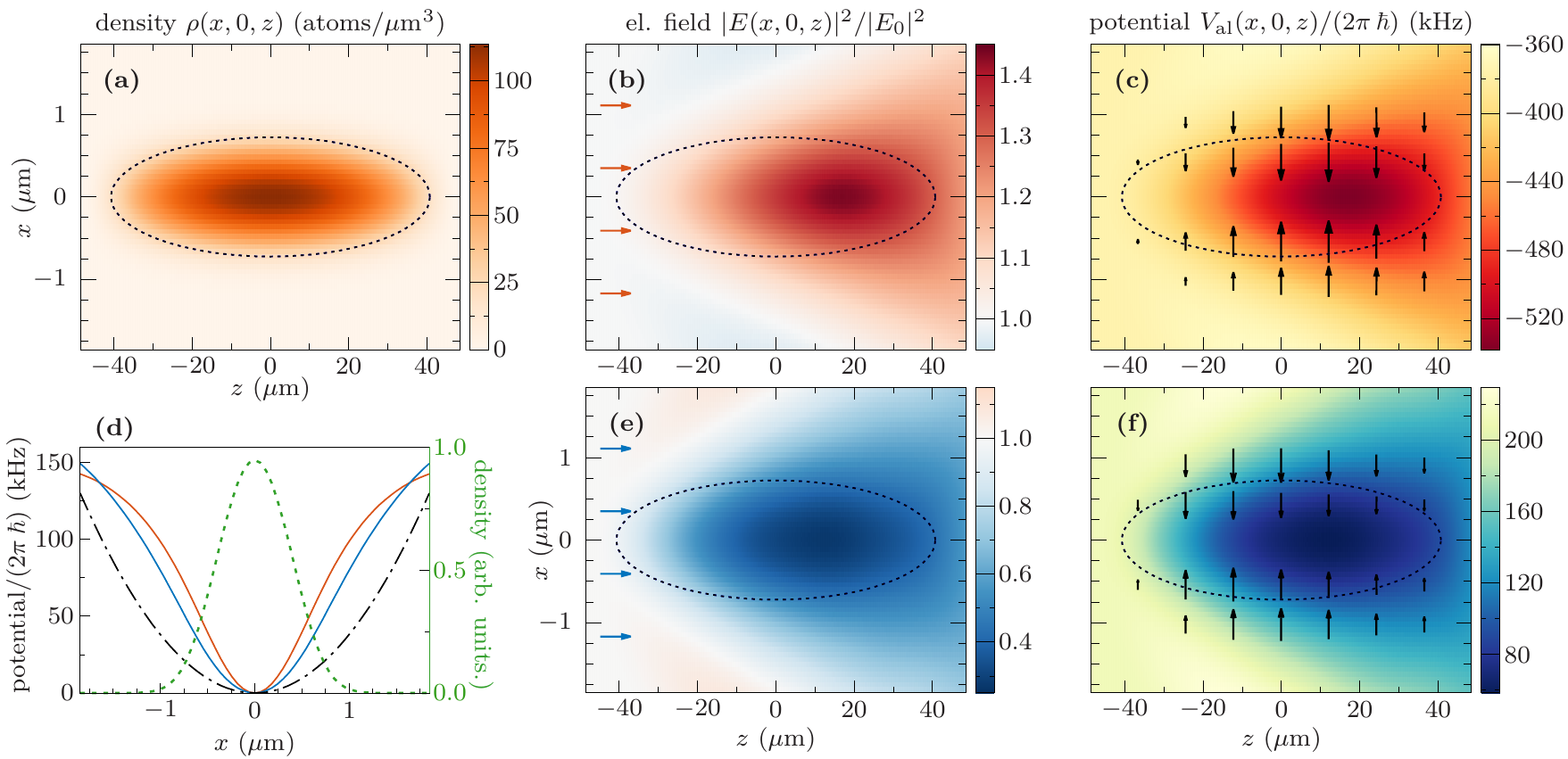}
    \caption{\label{fig:_ComparingPotentials}
    Simulated interaction between a 1d-BEC and red-/blue-detuned light. (a) Density distribution for $N=6200$ atoms after an expansion time of~$100\,\mathrm{\mu s}$. For red-detuned light ($\Delta=-392\Gamma$) this corresponds to a peak refractive index $\approx 1.002$. For blue-detuned light ($\Delta=+100\Gamma$) the refractive index is reduced to $\approx 0.993$. The black dotted ellipse indicates where the atom density drops to 10\% of its peak value. (b) The macroscopic electric field (solution to Eq.~\eqref{eq:_Helholtzequ}) from the interaction between a red-detuned plane laser wave (travelling from left to right, indicated by arrows) and the cloud of atoms (indicated by the dotted ellipse). We see that atoms act like a focusing lens. (c) Resulting atom-light potential (\cf Eq.~\eqref{eq:_Val}). The saturation is chosen as in Fig.~\ref{fig:_ExpansionAfterCloud_ExpAndTheory} with $s=319\times 10^{-6}$. (d) Radial potential at $y=z=0$ for red and blue-detuned case (red and blue solid line, respectively); the black, dash-dotted curve shows the original radial trapping potential $\hbar \omega_\perp x^2/(2 a_\perp^2)$ for comparison; the radial atomic density is indicated by the dotted green curve and measured by the right ordinate. The potentials are shifted such that $V(0)=0$. Note that the collective atom-light potentials are remarkably strong, comparable to the original magnetic trapping potential, but they have a very different shape. The LI-DD self-confining potential shows a depth equivalent to $\sim8\mu K$. (e) Same as (b), but for blue-detuned ($\Delta=+100\,\Gamma$) light. The atoms now act like a divergent lens. (f) Same as (c), atom-light potential for a saturation $s=708\times 10^{-6}$ and $\Delta=+100\,\Gamma$. In both figures (c) and (f) the highest potential energy corresponds to light interacting with a single atom. The spatial variation of the potential energy causes a compression for red as well as for blue detunings (force indicated by arrows).
    }
\end{figure*}

Fig.~\ref{fig:_ComparingPotentials} illustrates that for red-detuned light the 1d-BEC focuses the incoming laser beam as the refractive index increases with the particle density~\cite{mazets2000ground}. The resulting gradient in the light field pulls the atoms towards regions of higher particle density which leads to even stronger focusing of the light field. For high atomic densities, this self-focusing is counteracted by the repulsive $s$-wave scattering of the atoms and atom loss (discussed in Sec.~\ref{sec:_Results}).

For blue-detuned light, we have the opposite process leading to a similar effect: Here the refractive index drops below one and the cloud behaves like a divergent lens, pushing the light field away from the atoms. But for blue-detuned light, atoms are pulled towards regions of lower light intensity, such that they again accumulate in regions of high atomic density.

An essential difference between usual radiation forces and the LI-DD interaction is that the latter is an effective particle-particle interaction, mediated by scattered light. It does not trap atoms at a fixed position (for example, the focus of a laser beam), but draws them towards regions of maximum particle density. If the cloud is displaced, the LI-DD potential moves with it, as long as the atoms are (homogeneously) illuminated.

The arrows in Fig.~\ref{fig:_ComparingPotentials}c) and f) indicate the LI-DD force as a negative gradient of the respective atom-light potential. Here, due to the elongated geometry the force acts mainly in the radial direction. We also note that the forces are not symmetric around the center of the atomic cloud in the axial direction, a feature of their nonlocal behaviour (see also discussion in the context of Fig.~\ref{fig:_CompressionVsZ}).

\section{Results}\label{sec:_Results}

\begin{figure}
    \centering
    \includegraphics[width=\linewidth]{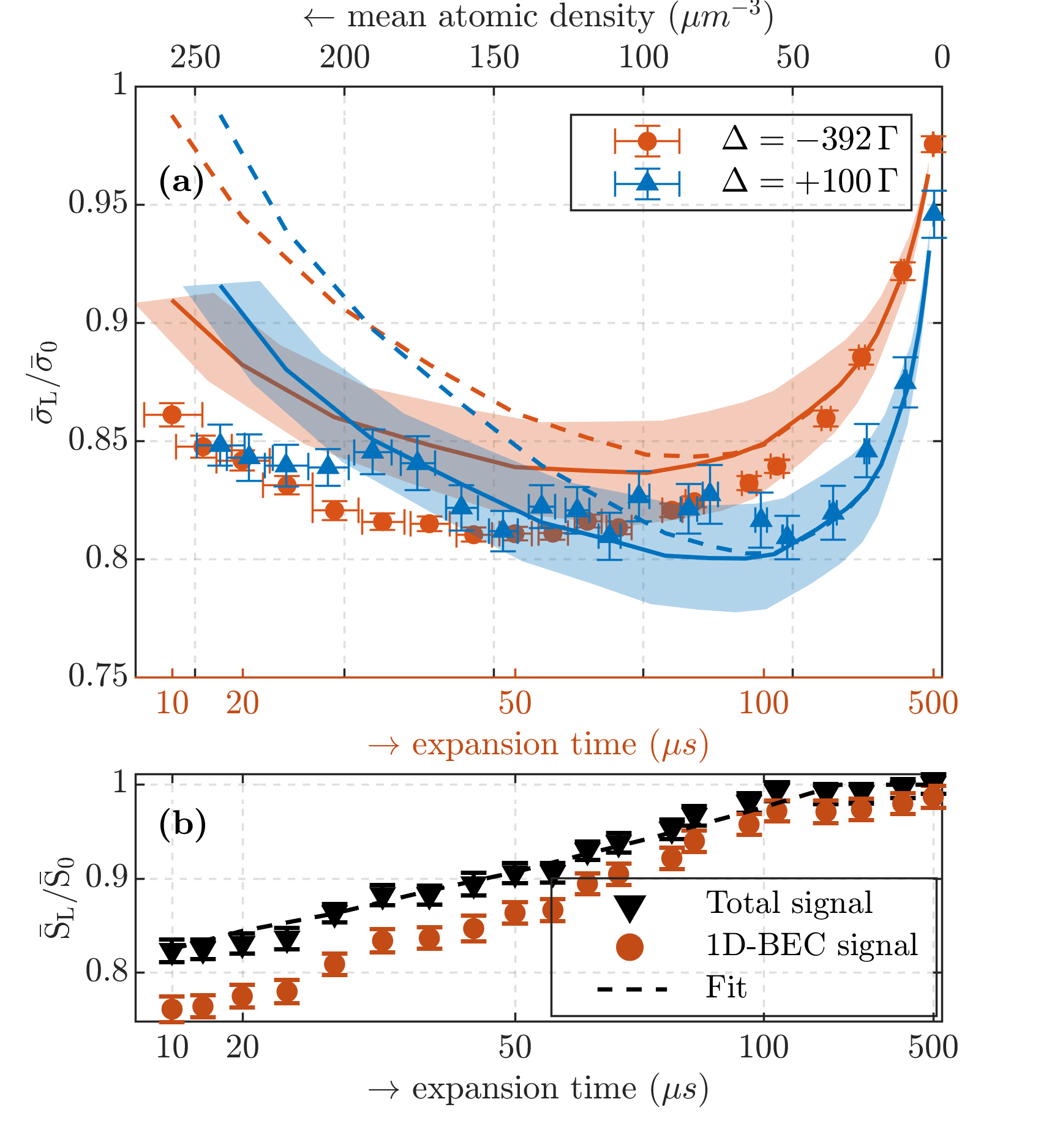}
    \caption{\label{fig:_ExpansionAfterCloud_ExpAndTheory}
    (a) Relative average transverse width $\bar{\sigma}_L/\bar{\sigma}_0$ of the 1d-BEC after ToF with ($\bar{\sigma}_L$) and without ($\bar{\sigma}_0$) LI-DD interaction at different expansion times (atomic densities). The blue triangles (red circles) depict results for a laser detuning of $\Delta = +100 \Gamma$ ($-392 \Gamma$) with a saturation parameter $s=708(20)\times 10^{-6}$ ($319(11)\times 10^{-6}$) and mean atom number $N=6600(130)$ (7450(60)). 
    The reduction of the relative mean transverse width is compared to the theoretical prediction with (solid lines) and without (dashed lines) considering the observed atomic loss, respectively (see text and dashed line in (b)). The shaded areas are theoretical predictions that account for a $\pm$10\% variation of the atom number and the saturation parameter.
    (b) Integrated fluorescence signal $\bar{S}_L$ after LI-DD interaction compared to the signal without laser pulse interaction $\bar{S}_0$ (for data set shown in (a) with $\Delta = -392 \Gamma$, losses for $\Delta = 100 \Gamma$ are similar).
    The black triangles show the integrated signal of the entire light sheet image, the red circles the signal in the bulk BEC, see also Fig.~\ref{fig:ROI}.
    For densities smaller than $50\, \mathrm{atoms/\mu m^3}$, only the density-independent single-photon scattering induced by the laser pulse is observed. For larger densities, additional mechanisms like super\-radiance and possibly light-assisted collisions cause atom loss of up to 25\% (red bullets).
   Error bars depict the standard error of the mean.}
\end{figure}

In our experimental data the most prominent effect of LI-DD interactions manifests itself in a compression, \ie a reduction of the transverse width of the atom cloud after time of flight. Fig.~\ref{fig:_ExpansionAfterCloud_ExpAndTheory}(a) shows the ratio $\bar{\sigma}_L/\bar{\sigma}_0$ of the average transverse width of the atom cloud with ($\bar{\sigma}_L$) and without ($\bar{\sigma}_0$) laser interaction for various mean atomic densities $\rho$, ranging from $260$ to $3\,\mathrm{atoms /\mu m^3}$ \footnote{The average atomic density is here defined as expectation value, $\langle\rho(t)\rangle = \frac{1}{N} \int \dd^3 r \rho(t,\br)^2$. For a Gaussian radial profile $\sim \exp[-r^2/(2 \sigma(t)^2)]$ and expanding width $\sigma(t) = \sigma(0) \sqrt{1+\omega_\perp^2 t^2}$ the mean density thus evolves as $\langle\rho(t) \rangle \approx \langle\rho(0) \rangle \sigma(0)^2/\sigma(t)^2$. This assumes that the longitunal density profile does not change during short expansion times.}.
We find the strongest compression for both red and blue-detuned light pulses at densities of $50$ to $150\,\mathrm{atoms/\mu m^3}$. For shorter expansion times corresponding to higher densities, the attractive LI-DD dynamics are still strongly influenced by the repulsive $s$-wave interaction between the atoms.

As explained in section~\ref{sec:_Theory}, the LI-DD force depends on the gradient of the atomic density and  linearly on the saturation parameter $s=\frac{I/I_\text{sat}}{1+4(\Delta/\Gamma)^2}$, with the saturation intensity $I_\text{sat}=3.12\,\mathrm{mW/cm^2}$, the decay rate $\Gamma = 2 \pi \times 6.067 \, \mathrm{MHz}$ and the detuning between the atomic resonance and the laser frequency $\Delta=\omega_0-\omega_L$ (see Eq.~\eqref{eq:_Val_linear}).
Note that for similar parameters we observe a stronger compression for red than for blue-detuned light fields, \cf also Fig.~\ref{fig:_ComparingPotentials}(d). This effect can be explained intuitively by the fact that atoms ``expel" blue detuned light fields from the regions of high atomic densities such that most particles are in regions of reduced intensity and weaker interaction.

\begin{figure}
    \centering
     \includegraphics[width=\linewidth]{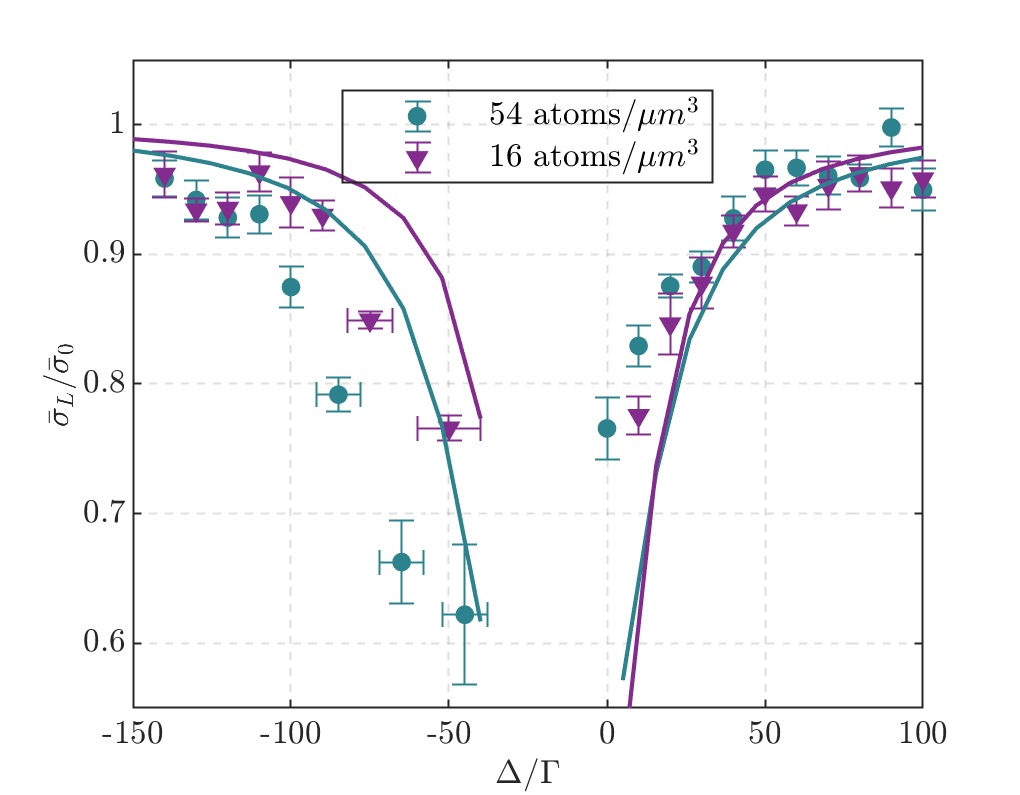}
    \caption{\label{fig:_CompressionVsDetuning}
   Relative average transverse width after illuminating the BEC for $5\,\mathrm{\mu s}$ (with beam intensity $I=11.54~\mathrm{mW/cm^2}$ or $s=150\times 10^{-6}$ at $\Delta=-100 \Gamma$) as a function of the detuning $\Delta$ after 100 $\mu s$ ($\corr 54\, \mathrm{atoms /\mu m^3}$, turquoise circles) and 200 $\mu s$ ($\corr 16\, \mathrm{atoms /\mu m^3}$, purple triangles) free expansion time. We observe a reduction of the transverse width for red and blue detunings while the maximum compression is red shifted from the bare atomic resonance at $\Delta=0$. 
   Close to resonance, we observe not only single-photon scattering, but also the onset of superradiance and additional atom loss (see text). These phenomena are not included in the numerical simulations (solid lines) which therefore only qualitatively match the data. The error bars depict the standard error of the mean. Due to the strong loss close to resonance, the data have been post-selected based on the remaining atom number in the 1d-BEC (\cf Fig.~\ref{fig:ROI}).
   }
\end{figure}

For mean atomic densities lower than $50\,\rm{atoms/\mu m}^3$ the compression due to LI-DD interactions is accompanied only by $\sim 3\%$ of single-photon scattering which is visible as a halo in the light sheet images (see Fig.~\ref{fig:_ExpansionAfterCloud_ExpAndTheory}(b) and appendix~\ref{sec:_measurement_of_saturation}). 
In this regime the experimental data are in good agreement with the numerical simulations based on the theoretical model discussed in Sec.~\ref{sec:_Theory} for both red ($\Delta = -392 \Gamma$) and blue ($\Delta = 100 \Gamma$) detunings.

However, for dense atomic samples illuminated by intense beams, we see two additional effects: Firstly, we observe onsets of superradiance through groups of atoms that are spatially separated from the initial 1d-BEC by $\pm 2\hbar k$ and $+4\hbar k$, see also Fig.~\ref{fig:ROI}(d). Atoms scattered to $-4\hbar k$ lie outside the light sheet imaging region.

Secondly we find that up to 18 \% of the atoms are missing from the light sheet images.
This indicates that the lost atoms are either not on resonance with the light sheet (\eg due to molecule formation) or they receive sufficient momentum to miss the imaging region after the long ToF. This special atom loss depends linearly on the mean atomic density $\rho$ (see black triangles in  Fig~\ref{fig:_ExpansionAfterCloud_ExpAndTheory}(b)), thus indicating a 2-body collision loss process. We attribute it to light-assisted collisions~\cite{fuhrmanek2012light}. Raman scattering into other magnetic or hyperfine sublevels of the $^{87}$Rb ground state can be ruled out, since these atoms would still appear in the light sheet images. 

The observed atom loss is not part of the theory model discussed in section~\ref{sec:_Theory}. In the numerical simulations we hence include a phenomenological term ~$-i \eta \abs{\psi}^2$ accounting for the losses through a free parameter~$\eta$ adjusted to reproduce the experimentally observed reduction of the atom number (\cf dashed line in Fig.~\ref{fig:_ExpansionAfterCloud_ExpAndTheory}(b)). As can be seen in Fig.~\ref{fig:_ExpansionAfterCloud_ExpAndTheory}(a), the inclusion of the observed atom losses in the simulations significantly reduces the gap between experiment and simulations for mean densities higher than $50\,\mathrm{atoms/\mu m^3}$.

We also observe asymmetric detuning-dependent effects close to resonance. Fig.~\ref{fig:_CompressionVsDetuning} shows the measured compression for a fixed laser beam intensity as a function of the detuning~$\Delta$. While for red detunings a clear compression is observable even for large $\abs{\Delta}$, we see LI-DD effects for blue detuning mainly close to resonance. This asymmetry is much stronger than what would be expected from the Clausius-Mossotti relation Eq.~\eqref{eq:_susceptibility} and points to the non-trivial wave propagation for the given refractive index distribution.

\begin{figure}
    \centering
    \includegraphics[width=\linewidth]{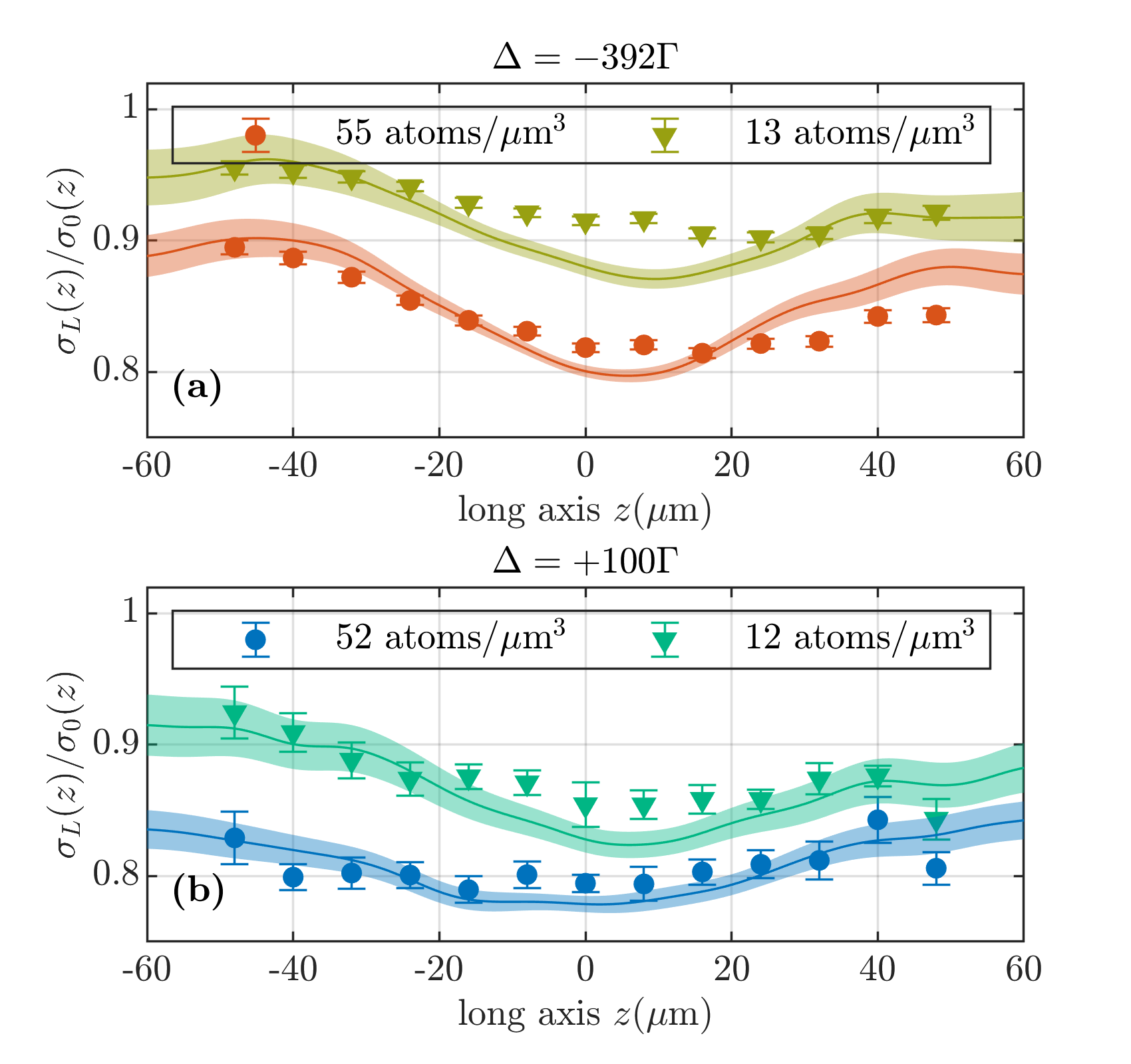}
    \caption{\label{fig:_CompressionVsZ}
    Relative transverse width $\sigma_L(z)/\sigma_0(z)$ as a function of the (axial) z-axis of the BEC (after ToF = $44\,\mathrm{ms}$) with $\sigma_L(z)$ and without $\sigma_0(z)$ interaction with a $5\mu s$ long laser pulse at a detuning of (a) $\Delta = -392 \Gamma$, atom number $N=7450(60)$, saturation $s=319(11)\times 10^{-6}$, and (b) $\Delta = +100 \Gamma$, $N=6600(130)$, $s=708(29)\times 10^{-6}$. The laser pulse propagates from left to right.
    The bullets (triangles) represent the measurements with the light pulse triggered after $105$ $(235)\, \mu \mathrm{s}$ expansion time. The error bars depict the standard error of the mean.
    The lines show the results of simulations including thermal phase and density noise for the condensate temperature~$T=135\,\mathrm{nK}$, see appendix~\ref{sec:_thermal_expansion}. The shaded areas depict the standard error of the mean obtained from 100 runs of the numerical simulation.  An asymmetric compression of the transverse width along the axial direction of the BEC is clearly observable for different mean atomic densities $\rho$ at the start of the LI-DD interaction. 
    This behavior arises due to the nonlocality of the LI-DD interaction.
    }
\end{figure}

In Fig.~\ref{fig:_CompressionVsZ} we show the radial compression as a function of the axial ($z$) position. For high atomic densities and red detunings the relative compression is strongest at $z>0$, \ie the maximum compression is not at the peak 1d-density (at $z=0$) of the quasi-condensate, but shifted in direction of the propagation axis of the laser beam. For blue detunings the maximum is shifted in opposite direction, \ie towards the beam source (\cf force arrows in Fig.~\ref{fig:_ComparingPotentials}(c) and (f)). This effect is less prominent for low atom density.

The observed features are in good agreement with numerical simulations including the finite temperature of the 1d-BEC (see appendix~\ref{sec:_thermal_expansion}). As discussed in Sec.~\ref{sec:_Theoretical_Intuition} the spatial dependence of the LI-DD force is determined by the (de-)focussing of the light field by the entire atomic cloud. The resulting inhomogeneous compression can thus be interpreted as a feature of the nonlocality of the LI-DD interaction.

\section{Theory}\label{sec:_Theory}

In a simplified model one can describe the BEC as a cloud of two-level atoms interacting with a classical electromagnetic field. Adiabatically eliminating the excited state one obtains an effective Gross-Pitaevskii equation for the ground-state atoms,~\cite{zobay2006spatial}
\begin{multline}\label{eq:_GPnew}
     i \hbar \partial_t \psi(\br) = 
    \left( -\frac{\hbar^2 \nabla^2}{2 m} + V_\text{T}(\br) + U \abs{\psi(\br)}^2 \right)\psi(\br)
    \\
     + \re \left[\frac{ \abs{\bd\cdot \mathbf{E}^{+}(\br)}^2}{\hbar (\Delta + i \Gamma/2)}\right]
    \psi(\br) \,.
\end{multline}
Here $V_\text{T}$ is the trapping potential which in our case is turned off before the interaction with the light field; $U=4\pi \hbar^2 a_s/m$ with the $s$-wave scattering length $a_s$ gives the mean-field interaction. The term in the second row of Eq.~\eqref{eq:_GPnew} describes the action of the light field on the atoms and will later be called atom-light potential~$V_\text{al}(\br, \psi)$.

The level of approximation of the electric field~$\mathbf{E}(\br) = \mathbf{E}^+(\br) e^{-i \omega_L t} + \mathrm{c.c.}$ determines which effects are included in the model: for example, assuming only a pure laser field, $\mathbf{E}^+(\br) \rightarrow \mathbf{E}_\text{L} e^{i \bk_\text{L}\cdot \br}$, gives an ensemble where each atom independently interacts with the laser.

Effective interactions between the atoms arise when we instead use the macroscopic electric field, $\mathbf{E}^+ \rightarrow \mathbf{E}_\text{m}^+$, fulfilling the inhomogeneous wave equation
\begin{equation}\label{eq:_waveequ}
    \nabla\times\left(\nabla \times \mathbf{E}_\text{m}^\pm \right) + \tfrac{1}{c^2} \partial_t^2 \mathbf{E}_\text{m}^\pm = - \tfrac{1}{\varepsilon_0 c^2}\partial_t^2 \mathbf{P}^\pm \,,
\end{equation}
with the polarization density $\mathbf{P}^+(\br)=\varepsilon_0 \chi(\br) \mathbf{E}_\text{m}^+(\br)$.
Unfortunately most geometries do not allow for a simple analytic solution of Eq.~\eqref{eq:_waveequ}, but with some simplifications it can be used to describe, for example, superradiance~\footnote{Note that describing superradiance using classical electromagnetic fields requires a sufficiently noisy matter field~$\psi$ to simulate the quantum fluctuations (spontaneous emission) required to seed superradiant growth~\cite{moore1999theory,zobay2006spatial,uys2008cooperative}.} in perpendicularly pumped BECs~\cite{zobay2006spatial}.
Using the Green's tensor~\cite{novotnyhecht2012principles,morice1995refractive}
\begin{multline}\label{eq:_Greenstensor}
	\mathbb{G}(\br) = 
	\frac{1}{4 \pi \varepsilon_0} \frac{e^{i k r}}{r^3}
				\Big[ \id \left( k^2 r^2 + i k r  - 1 \right)
				\\
				- \frac{\br \otimes \br}{r^2} \left( k^2 r^2 + 3 i k r - 3  \right) \Big]
			- \frac{1}{3 \varepsilon_0} \delta(\br) \id
\end{multline}
one can formally solve Eq.~\eqref{eq:_waveequ} for an incoming plane wave field $\mathbf{E}_\text{L} e^{i \bk_\text{L}\cdot\br}$ to
\begin{equation}\label{EQ:_Efield_solution}
	\mathbf{E}_\text{m}^+(\br) = \mathbf{E}_\text{L} e^{i \bk_\text{L}\cdot\br} + \int \dd^3 r'\, \mathbb{G}(\br-\br') \mathbf{P}^+(\br')\,.
\end{equation}
Using a single-scattering (first Born) approximation we write~$\mathbf{P}^+ \approx - \bd (\bd\cdot \mathbf{E}_\text{L})e^{i \bk_\text{L}\cdot \br} \abs{\psi}^2 /[
\hbar (\Delta + i \Gamma/2)]$ in Eq.~\eqref{EQ:_Efield_solution} and use the result in Eq.~\eqref{eq:_GPnew}. Neglecting products of Green's tensors one then arrives at an effective atom-light potential~\cite{zhang1994quantum,odell2000bose,giovanazzi2001one,giovanazzi2002density},
\begin{multline}\label{eq:_Val_linear}
    V_\text{al}(\br, \psi) \approx
    \re\left[ \frac{ \abs{\bd\cdot \mathbf{E_\text{L}}}^2}{\hbar (\Delta + i \Gamma/2)} \right] \bigg( 1 -
    \\
     - \frac{ 3 \Gamma}{2}  \re \bigg[
    \int \dd^3 r'\, \frac{ \mathcal{G}(\br-\br')}{(\Delta + i \Gamma/2)} \abs{\psi(\br')}^2  \bigg] \bigg)\,,
\end{multline}
with a dimensionless interaction kernel $\mathcal{G}$ defined via
\begin{equation}
    \bd \cdot\left( \left[ \mathbb{G}(\br) + \tfrac{\delta(\br)}{3\varepsilon_0}\right]\cdot \bd\right) e^{-i \bk_L\cdot\br}  = \frac{k_0^3 \abs{\bd}^2}{4 \pi \varepsilon_0} \mathcal{G}(\br)\,.
\end{equation}
Here $k_0=\omega_0/c$ and $\Gamma = k_0^3 \abs{\bd}^2/(3 \pi \varepsilon_0 \hbar)$ is the single-atom decay rate. The term $\delta(\br)/(3 \varepsilon_0)$ was added to cancel the self-interaction from Eq.~\eqref{eq:_Greenstensor} \footnote{This is equivalent to excluding a small spherical volume around $\br'=\br$ from the integration~\cite{krutitsky1999local,born1999principles}, see also Ref.~\cite{ruostekoski1997lorentz} for a discussion on the the self-interaction and the Lorentz-Lorenz model.}.

The form given in Eq.~\eqref{eq:_Val_linear} clearly separates the interaction between individual atoms and the incoming laser field from the effective LI-DD potential. It is also visible that LI-DD couplings are generally proportional to $\Delta^{-2}$ and scale linearly  with the light intensity or atom number. But as we discuss in appendix~\ref{sec:_estimates_approx_dipole_dipole}, such a truncated model is only valid if the effects of multiple scattering between the atoms and the light field can be neglected, see Fig.~\ref{fig:_ComparingTheories}.

To include effects of multiple scattering, which is significant at the atom densities used in our experiment, we use a different model in the present work. Since atoms actually interact with the \emph{local} field we set $\mathbf{E}^+ \rightarrow \mathbf{E}_\text{loc}^+ =\mathbf{E}_\text{m}^+ - \mathbf{P}^+/(3 \varepsilon_0)$ in Eq.~\eqref{eq:_GPnew}. While $\mathbf{E}_\text{m}^+$ is still the macroscopic field solving Eq.~\eqref{eq:_waveequ}, the polarization in Eqs.~\eqref{eq:_waveequ} or~\eqref{EQ:_Efield_solution} is then proportional to the local field, $\mathbf{P}^+ = - \bd (\bd\cdot \mathbf{E}_\text{loc}^+) \abs{\psi}^2 /[\hbar (\Delta + i \Gamma/2)]$~\cite{born1999principles}.

In the current experiment the incoming laser field is $\sigma^-$- polarized to drive the transition between the $F=1$, $m_F=-1$ and $F'=2$, $m_F'=-2$ states. 
Due to the elongated geometry of the cloud, light scattering in the forward and backward direction dominates the interaction. Since this scattering is polarization conserving we assume that the macroscopic field is also $\sigma^-$-polarized and parallel to the dipole vector $\bd= \abs{\bd} \be_d$. This allows us to neglect scattering to other internal states and leads directly to the scalar atomic polarizability, $\alpha = - \frac{ \abs{\bd}^2 }{\hbar \left( \Delta + i \Gamma/2\right)}$, and the susceptibility fulfilling the Clausius-Mossotti or, equivalently, Lorentz-Lorenz relation,
\begin{equation}\label{eq:_susceptibility}
	\chi(\br) = \frac{\abs{\psi(\br)}^2 \alpha/\varepsilon_0}{1-\abs{\psi(\br)}^2 \alpha/(3 \varepsilon_0)}
	\,.
\end{equation}
Approximating the incoming laser field as a plane wave of amplitude $\mathcal{E}_\text{L}$ we can define the macroscopic field as $\mathbf{E}_\text{m}(\br) :=\mathcal{E}_\text{L} \Theta(\br) \be_d$. The mode function $\Theta(\br)$ is then obtained by solving the Helmholtz equation
\begin{equation}\label{eq:_Helholtzequ}
	\left(\nabla^2  + k_L^2\left[1 + \chi(\br) \right] \right) \Theta(\br) = 0
	\,.
\end{equation}

Introducing the Rabi frequency for the incoming field, $\hbar \Omega_\text{L}/2 = -\abs{\bd} \mathcal{E}_\text{L}$, and the saturation for a single atom, $s(\Delta) = (\abs{\Omega_L}^2/2)/(\Delta^2 + \Gamma^2/4)$, we can rewrite the atom-light potential $V_\text{al}(\br) = \re[ \alpha] \abs{\mathbf{E}_\text{loc}(\br)}^2$ as
\begin{equation}\label{eq:_Val}
	V_\text{al}(\br) = 
	    - \frac{ \hbar s(\Delta)}{2}  \frac{\Delta}{\abs{1 + \alpha \abs{\psi(\br)}^2/(3 \varepsilon_0)}^2} \abs{\Theta(\br)}^2
		\,.
\end{equation}
In this form there is no obvious distinction between atoms coupling to the free laser field and the effective light-induced dipole-dipole interaction. The atoms collectively reshape the electromagnetic field and each atom then interacts with the resulting local field. We thus see that the particle density enters the potential both through the prefactor, but also indirectly (nonlocally) via the solution of Eq.~\eqref{eq:_Helholtzequ} with the susceptibility given in Eq~\eqref{eq:_susceptibility}.

Numerical simulations of Eq.~\eqref{eq:_Helholtzequ} and the resulting atom-light potential for typical experimental parameters are shown in Fig.~\ref{fig:_ComparingPotentials}. The potential energy~$V_\text{al}(\br)$ has its minimum approximately at the center of the cloud and a steep radial gradient. Outside the cloud, the potential reaches the constant value for a single atom in a plane wave ($\abs{\Theta} \rightarrow 1$, $\abs{\psi}^2\rightarrow 0$).

The short discussion here is mostly a qualitative explanation. Of course, the effective atom-light potentials given in Eqs.~\eqref{eq:_Val_linear} and~\eqref{eq:_Val} can be derived rigorously~\cite{thirunamachandran1980intermolecular,craig1984molecular,morice1995refractive,zhang1994quantum,krutitsky1999local,javanainen1995off,ruostekoski1997quantum,ruostekoski1997lorentz}. Such a derivation shows that higher-order density correlations have to be neglected in order to obtain the classical susceptibility from Eq.~\eqref{eq:_susceptibility} \cite{morice1995refractive,ruostekoski1997lorentz,guerin2017light}. Numerical simulations based on classical dipoles also show that the Lorentz-Lorenz (Clausius-Mossotti) model holds for thermal samples where the effect of correlation functions is smeared out~\cite{javanainen2014shifts,jenkins2016collective}.

For the low densities of the present experiment, the possible corrections to the Lorentz-Lorenz relation are expected to be small~\cite{guerin2017light}. Indeed, $\abs{\psi(\br)}^2 \alpha/\varepsilon_0  \sim 10^{-2} $ for typical densities and detunings. Additionally, in a quasi-condensate the long-range order breaks down due to thermal phase and density noise \cite{popov1987functional,mora2003extension} (see also appendix~\ref{sec:_thermal_expansion}). However, higher order correlations might point the way to resolve the discrepancy between experiment and simulation observed for higher densities in Fig.~\ref{fig:_ExpansionAfterCloud_ExpAndTheory}.

\subsection{Numerical Modelling}\label{sec:_NumericalModel}

\begin{figure}
    \includegraphics[width=\linewidth]{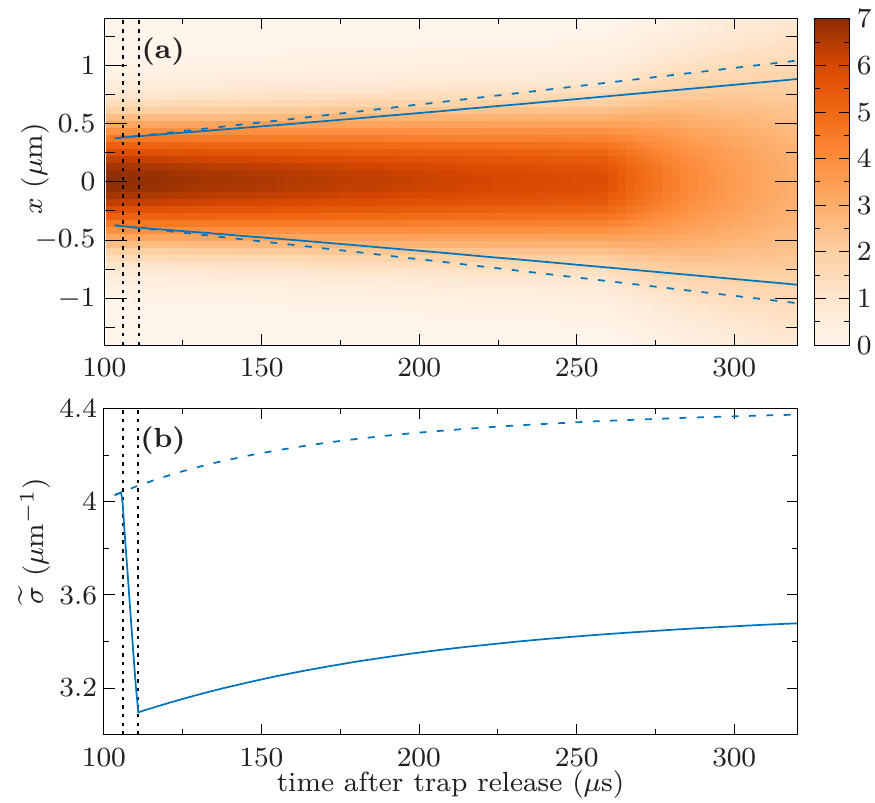}
    \caption{\label{fig:_ExpandingCloud}
    Simulated expansion of a cloud of $N=6600$ atoms with (solid lines) and without (dashed lines) the interaction with a spatially homogeneous laser pulse. The pulse with a saturation $s=708\times 10^{-6}$ and a detuning $\Delta=+100 \Gamma$ starts after an expansion time $t_\mathrm{exp}=105\,\mathrm{\mu s}$ and lasts for $5~\mathrm{\mu s}$. In the trap, at $t=0$, the cloud had a Gaussian width $a_\perp\approx 200\,\mathrm{nm}$. (a) Evolution of the average radial density $\rho_\perp(t,x) = \int \dd y \int \dd z \,\abs{\psi(t,x,y,z)}^2$ (in $1000\,\mathrm{atoms/\mu m}$, see color scale on the right), the lines indicate the width~$\sigma$ of a fitted Gaussian~$\sim \exp[-x^2/(2\sigma^2)]$. (b) Transverse width of the Fourier transformed wave functions, $\widetilde{\rho}_\perp(t,q_x) = \int \dd q_y \int \dd q_z \,\abs{\widetilde{\psi}(t,q_x,q_y,q_z)}^2 \sim \exp[-q_x^2/(2\widetilde{\sigma}^2)]$. The dashed vertical lines indicate the interaction time with the light pulse. We see that the width of the cloud does not change during the pulse, but the width of its Fourier transform (its momentum distribution) does. During the interaction the wave function accumulates a position dependent phase, which explains why the width in real and Fourier space are reduced despite the otherwise free expansion.}
\end{figure}

Our numerical procedure starts with calculating the ground state of the 1d-BEC for a given atom number in the trap described in section~\ref{sec:S1-Experiment basics} using a split-step method with imaginary time propagation. This state then evolves without a trap in three dimensions for up to~$400\,\mathrm{\mu s}$ (\cf expansion time in Fig.~\ref{fig:SETUP}(b)). During this time the average width of the 1d-BEC expands as $\sigma(t) = \sigma(0) \sqrt{1 + (\omega_\perp t)^2}$~\cite{castin1996bose}.

Subsequently, light of a given saturation $s$ and detuning $\Delta$ is switched on for~$5~\mathrm{\mu s}$. Calculating the atom-light potential~\eqref{eq:_Val} then requires solving the Helmholtz equation~\eqref{eq:_Helholtzequ} for an inhomogeneous refractive index $\sqrt{1+\chi(\br)}$. This is done using the \emph{Wavesim} package for \emph{Matlab} by Osnabrugge et al.~\cite{osnabrugge2016convergent,osnabrugge2021ultra,WavesimSource}, which essentially gives an iterative solution to the integral equation for the electric field~\eqref{EQ:_Efield_solution}.

During the interaction time we thus use a split-step method to simulate the three dimensional evolution of Eq.~\eqref{eq:_GPnew} with the atom-light potential given in Eq.~\eqref{eq:_Val}. The dimensionless function $\Theta(\br)$ is obtained using the \emph{Wavesim} package and is repeatedly updated during the atom-light interaction to account for changes in the atomic density.

Once the light is switched off, the wave function evolves without external potentials for another~$420\,\mathrm{\mu s}$. At this stage, the cloud is dilute enough to ignore $s$-wave scattering and any further expansion is purely ballistic. The transverse width of the Fourier transform of this expanded state is thus used as a theory model for the experimentally measured width after ToF (\cf Fig.~\ref{fig:_ExpandingCloud}).

To calculate the $z$-dependent compression shown in Fig.~\ref{fig:_CompressionVsZ} we simulate the full expansion up to $44\,\mathrm{ms}$ of ToF. Due to the long expansion time we further include thermal phase and density noise of the condensate during the ballistic expansion, details are given in appendix~\ref{sec:_thermal_expansion}.

Up to this point, the entire theoretical model is determined purely by the experimental settings and has no adjustable parameters. The model is sufficient to describe the collective dispersive coupling between the atoms and the light field, provided the densities are low enough to permit the use of the effective susceptibility in Eq.~\eqref{eq:_susceptibility}~\cite{guerin2017light}. Superradiant emission into $\pm 2 \hbar k$ momentum states could be included by adding sufficient seed noise to the initial ground state~\cite{uys2008cooperative}.

As mentioned in Sec.~\ref{sec:_Results}, the electrodynamical model of Sec.~\ref{sec:_Theory} has to be supplemented to include the experimentally observed atom losses, which we attribute mainly to light-assisted collisions. Such losses decrease the pressure due to $s$-wave scattering and thus reduce the expansion rate of the cloud. Therefore this leads to a larger relative compression as compared to the expansion without losses.

We model these losses by adding a density-dependent phenomenological loss term $V_\text{loss} = - i \eta\, \abs{\psi}^2$ to Eq.~\eqref{eq:_GPnew}, the only free parameter of our model. This allows us to estimate the effect of particle loss on the radial compression shown in Fig.~\ref{fig:_ExpansionAfterCloud_ExpAndTheory}. 

At high densities the repulsive $s$-wave interaction between the atoms counteracts the effects of the LI-DD interactions. Therefore, atom losses are the main cause for the observed compression at short expansion times, but play only a negligible role after long expansion times. At low densities the observed reduction of the transverse width is mainly an effect of LI-DD interactions.

\section{Conclusion}\label{sec:_Conclusion}

In summary, we demonstrated that a freely expanding 1d-BEC illuminated homogeneously along its axial direction experiences a strong, compressing radial force. Our numerical simulations show that a conceptually simple theoretical model using a mean-field approach, which does however require significant computational resources, successfully describes the experimental results for densities $\lesssim 50\, \mathrm{atoms/\mu m^3}$. At higher densities we observe strong atom loss, which we attribute to light-assisted collisions~\cite{fuhrmanek2012light}. The discrepancy between simulation and experiment at higher density is mainly caused by this loss.

The observed spatial dependence of the compression can be explained as a result of focusing (or defocusing, depending on the sign of the detuning) of the laser light in the atomic cloud, which is a manifestation of the nonlocality of the LI-DD interaction in our experimental setup~\cite{mazets2000ground}. Illuminating the medium by many laser beams shining from different directions~\cite{odell2000bose,ostermann2016spontaneous} or by light transmitted through a waveguide~\cite{shahmoon2016highly,griesser2013light,holzmann2015collective} would show further non-trivial aspects of the LI-DD interaction's nonlocality, for example their ability to establish long-range correlations between ultracold atoms.

Spatially varying level shifts associated with collective atom-light interactions may not only provide an additional noise source for high-precision atomic quantum sensors, 
but also enable interesting options to be used as a tool.
LI-DD interactions can give rise to an attractive $1/r$ inter-atomic potential, which allows to simulate and study “gravitational-like” interaction between highly delocalised (ultracold) quantum particles \cite{odell2000bose}. This interaction is in principle tunable over several orders of magnitude and could enable simulations of astrophysical scenarios \cite{artemiev2004electromagnetically} performed in quantum-controlled setups.

Since the LI-DD force is density dependent, expanding atomic clouds will be directed towards regions of higher density, a property which could be used for delta kick collimation \cite{ammann1997delta,deppner2021collective}. 
It is interesting to note that the LI-DD potential shows similar confinement strengths as used for trapping ultracold atomic gasses \cite{folman2002microscopic,borselli2021two}.

In addition, the LI-DD interaction is an effective, compressing atom-atom potential with a shape self-consistently determined by the atom cloud which may open complementary techniques for coherent collective manipulation of atomic ensembles, \eg by a pulsed optical grating~\cite{hornberger2012colloquium}. 

\section*{Acknowledgements}
We thank Helmut Ritsch for helpful remarks and discussions. P.H., M.S. and D.R. acknowledge the hospitality of the Erwin Schr\"odinger Institute in the framework of their ``Research in Teams" project. This research was supported by the Austrian Science Fund (FWF): through the START grant of P.H. (Y1121) and the DFG/FWF Collaborative Research Centre `SFB 1225 (ISOQUANT)'. Furthermore P.H. acknowledges funding by the 'Erwin Schr{\"o}dinger  fellowship': (FWF: J3680), and D.R. acknowledges funding by the Marie Sk\l{}odowska-Curie Action IF program -- Project-Name ``Phononic Quantum Sensors for Gravity" (PhoQuS-G) -- Grant-Number 832250". T.Z. acknowledges support from EU's Horizon 2020 program under Marie Sk\l{}odowska-Curie Grant No. 765267(QuSCo). I.M. acknowledges the support by the Wiener Wissenschafts- und Technologiefonds (WWTF) via project No. MA16-066 (SEQUEX). S.E. acknowledges support through an ESQ (Erwin Schr{\"o}dinger Center for Quantum Science and Technology) fellowship funded through the European Union’s Horizon 2020 research and innovation program under Marie Sk\l{}odowska-Curie Grant Agreement No 801110. This project reflects only the author’s view, the EU Agency is not responsible for any use that may be made of the information it contains. ESQ has received funding from the Austrian Federal Ministry of Education, Science and Research (BMBWF).

\clearpage
\appendix
\renewcommand\thefigure{\thesection\arabic{figure}}

\section{Estimates based on approximate dipole-dipole potential}\label{sec:_estimates_approx_dipole_dipole}
In Eq.~\eqref{eq:_Val_linear} we give a simplified potential for the atom-light interaction, which is based on a single-scattering (first Born) approximation for light propagation. Expressing the potential this way has the benefit that it clearly separates the coupling between the atoms and the incoming laser field from the effective dipole-dipole interaction, which is second order in the detuning. The nonlocal character of LI-DD interactions is visible from the convolution integral which shows that the interaction at point~$\mathbf{r}$ depends on the density of the entire sample. This potential has been used to predict compression and density modulations in elongated BECs and effective $1/r$ potentials in specific geometries~\cite{giovanazzi2001one,giovanazzi2002density,odell2000bose}.

However it is clear that the perturbative approach for light propagation can only be valid at low particle densities where multiple scattering events can be neglected. At higher densities one has to switch to numerical simulations either in particle models~\cite{javanainen2014shifts,andreoli2021maximum} or solving the wave equation, as we did.

\begin{figure}[b]
    \centering
    \includegraphics[width=\linewidth]{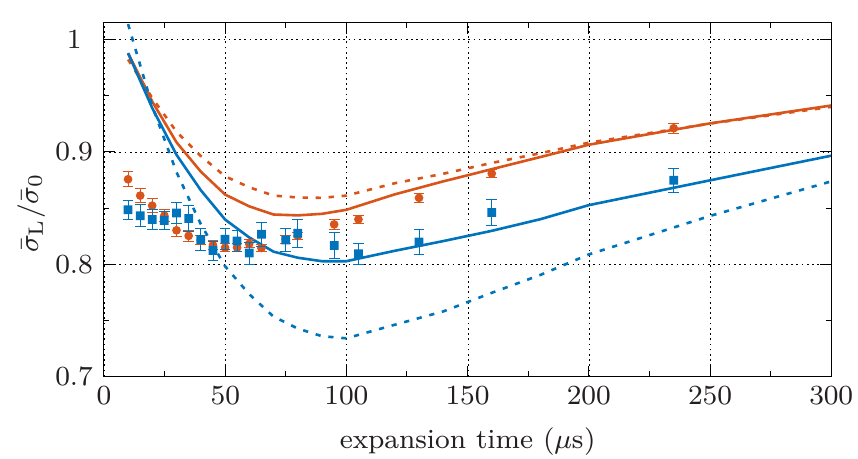}
    \caption{\label{fig:_ComparingTheories}
    Simulations of the relative average transverse width, $\bar{\sigma}_\text{L}/\bar{\sigma}_0$, of a BEC illuminated after different expansion times. The solid lines show the compression calculated with the full atom-light potential from Eq.~\eqref{eq:_Val} (used also for Fig.~\ref{fig:_ExpansionAfterCloud_ExpAndTheory}) while the dashed lines use the lowest order approximation given in Eq.~\eqref{eq:_Val_linear}. The blue (red) lines are for detuning $\Delta =  +100\, \Gamma$ ($\Delta=-392\, \Gamma$), atom number $N=6600$ $(7450)$ and saturation $s=708\times 10^{-6}$ $(319\times 10^{-6})$. For expansion times $>100\,\mathrm{\mu s}$ we see that the full model is in good agreement with the experimental data (indicated by circles and squares) while the approximate potential~\eqref{eq:_Val_linear} gives a misleading prediction, especially for the blue-detuned case.}
\end{figure}

In Fig.~\ref{fig:_ComparingTheories} we show simulations based on the single-scattering potential~\eqref{eq:_Val_linear} (dashed lines) for the data presented also in Fig.~\ref{fig:_ExpansionAfterCloud_ExpAndTheory}. As expected, this approximation performs less favourable as compared to simulations using the more evolved approach given in Eq.~\eqref{eq:_Val} (solid lines). The simulations shown in Fig.~\ref{fig:_ComparingTheories} do not account for particle losses and therefore are at odds with the data for short expansion times (higher densities).

\section{Inclusion of thermal expansion}\label{sec:_thermal_expansion}
\begin{figure}
    \centering
    \includegraphics[width=\linewidth]{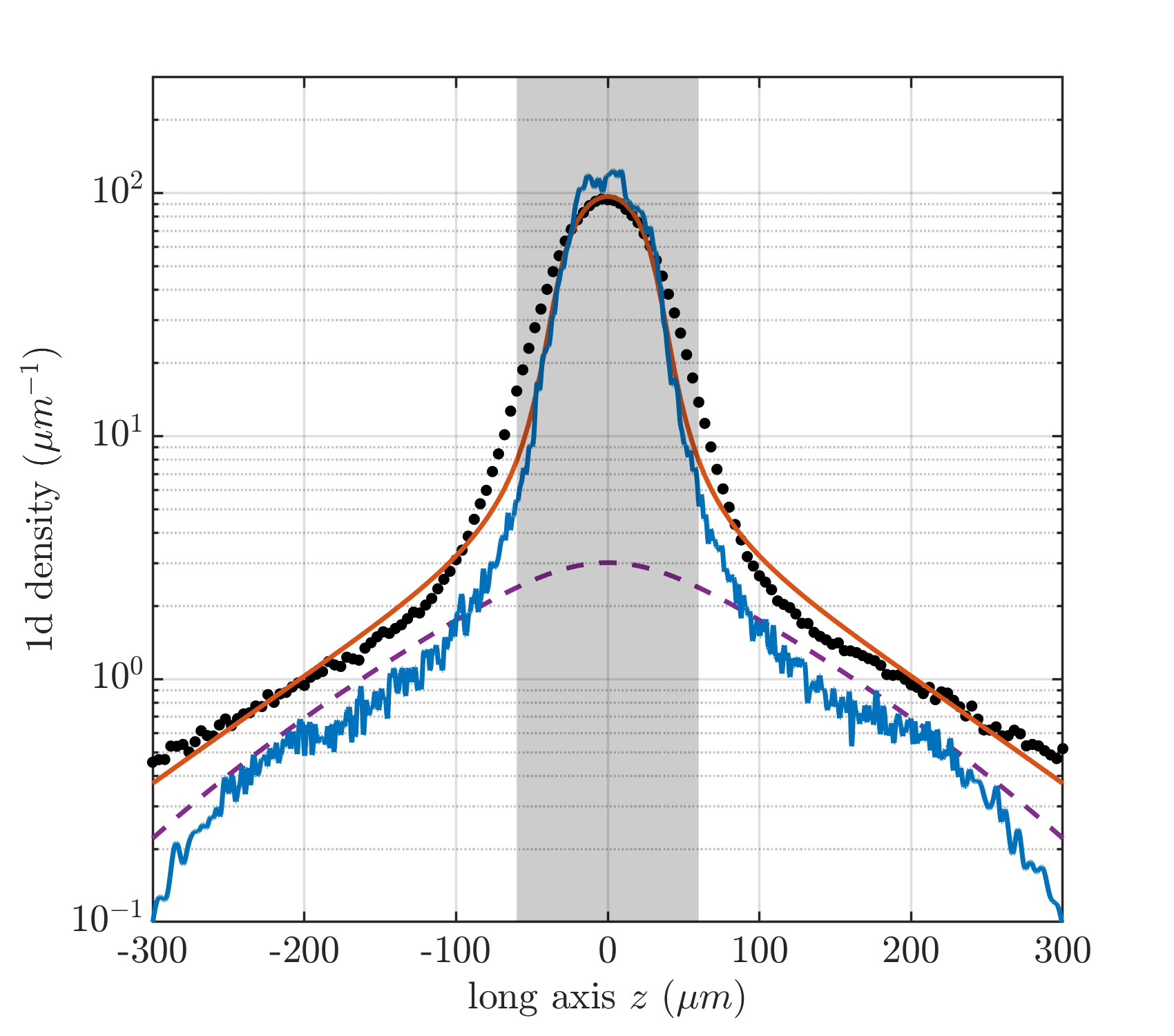}
    \caption{\label{fig:_Thermal_qBEC}
 Measured 1d-density after $44\,\mathrm{ms}$ ToF for $N=7450(60)$~atoms (black dots, average over 10 repeats) compared to the simulated 1d-density including phase and amplitude noise for $N=7450\, \rm{atoms}$ and $T=135\,\rm{nK}$ after 44 ms ToF, averaged over 100 repeats (blue line, see text). The red line shows a bimodal fit to the data yielding a temperature of $T=135(10)\, \rm{nK}$. The density of the condensed fraction is modeled with a Yang-Yang profile, the thermal fraction containing $\sim 10\%$ of the atoms by a Bose function (dashed purple line)~\cite{davis2012yytherm}. The grey shaded area depicts the plot range in Fig.~\ref{fig:_CompressionVsZ} of the main text.}
\end{figure}

The theory presented in Sec.~\ref{sec:_Theory} and the simulations used for Figs.~\ref{fig:_ComparingPotentials}, \ref{fig:_ExpansionAfterCloud_ExpAndTheory}, \ref{fig:_CompressionVsDetuning}, and~\ref{fig:_ExpandingCloud} assume a perfect condensate at temperature $T=0$. However, in the present experiment with a large atom number in a 1d quasi-condensate, the experimentally estimated temperature is $T=135(10)\,\mathrm{nK}$ (see Fig.~\ref{fig:_Thermal_qBEC}).

The non-zero temperature of a 1d-BEC manifests itself in a phase and density-noise along the axial direction, $\delta \phi(z)$ and $\delta \rho(z)$. For short expansion times (approximately less than $1\,\mathrm{ms}$) this noise has little effect on the evolution of the 1d-BEC, especially on the radial dynamics which we are mostly interested in.

But the phase and density noise associated with the finite temperature become relevant for the long-time expansion along the longitudinal direction. This is important when we want to compare simulation and experiment for the $z$-dependent compression $\sigma_\text{L}(z)/\sigma_0(z)$ shown in Fig.~\ref{fig:_CompressionVsZ}. The experimental data is taken after a time of flight $t_\text{ToF}=44\,\mathrm{ms}$.

To calculate the long-time expansion at $T>0$ shown in Fig.~\ref{fig:_CompressionVsZ} we therefore apply the following numerical procedure:
\begin{enumerate}
    \item\label{NumProcThermalExpansion_itemNoise} Calculate the in-trap phase and amplitude noise $\delta \phi(z)$ and $\delta \rho\text(z)$ for the given temperature and trap parameters. For an in-trap 1d-density $\rho_{1D}(0,z) = \iint \dd x\,\dd y\, \abs{\psi(0,x,y,z)}^2$ they are given by the Ornstein-Uhlenbeck processes~\cite{gillespie1996exact,stimming2010fluctuations,beck2018nonperturbative},
    \begin{align}
        \tfrac{\dd}{\dd z} \delta\phi(z) &= -\delta\phi(z)/\Lambda_\phi + g_\phi(z)
        \\
        \tfrac{\dd}{\dd z} \delta\rho(z) &= -\delta\rho(z)/\Lambda_\rho + g_\rho(z)
    \end{align}
    with $g_\phi$, $g_\rho$ being Gaussian random forces with zero mean while $\langle g_\phi(z) g_\phi(z') \rangle = \kappa_T(z) \delta(z-z')$ and $\langle g_\rho(z) g_\rho(z') \rangle = 4 \kappa_T(z) \rho_{1D}^2(0,z) \delta(z-z')$ with $\kappa_T(z) = m \mathrm{k_B} T /(\hbar^2 \rho_{1D}(0,z))$. The relaxation lengths are $\Lambda_\rho(z) = \frac{\hbar}{2} \left(2 m \hbar \omega_\perp a_s \rho_{1D}(0,z) \right)^{-1/2}$ and $\Lambda_\phi \rightarrow \infty$ for our setup.
    \item\label{NumProcThermalExpansion_itemBasicExpansion} Simulate the expansion after trap release, the interaction with the light field and the subsequent expansion for another $\sim 420\,\mathrm{\mu s}$ for $T=0$ as described in Sec.~\ref{sec:_NumericalModel}. After this second expansion time the effects of $s$-wave scattering can be neglected and all further expansion is ballistic.
    \item Step~\ref{NumProcThermalExpansion_itemBasicExpansion} returns a wave function~$\psi(t_1,x,y,z)$ on a 3d grid with an approximately Gaussian transverse density of width $\sigma(t_1,z)$. During the remaining time of flight the radial width expands from $\sigma(t_1,z) \approx 2\,\mathrm{\mu m}$ to $\sigma(t_\text{ToF},z) \approx 180\,\mathrm{\mu m}$. To reduce the memory load we thus switch to a radially symmetric wave function with the radial grid chosen such that both $\psi(t_1,r,z)$ and $\psi(t_\text{ToF},r,z)$ are well resolved.
    \item\label{NumProcThermalExpansion_AddNoise} Add the phase- and density noise calculated in step~\ref{NumProcThermalExpansion_itemNoise} via
    \begin{multline}
        \breve{\psi}(t_1,r,z) = \sqrt{\abs{\psi(t_1,r,z)}^2 + f_\perp(r,z)\,\delta \rho(z) } \\
        \times \exp\left[i \phi(t_1,r,z) + i \delta \phi(z)\right] \,,
    \end{multline}
    with $\phi(t_1,r,z) = \arg(\psi(t_1,r,z))$ and a radial distribution $f_\perp(r,z) \simeq \exp(-r^2/[2 \sigma^2(t_1,z)])$.
    \item The ballistic expansion of the wave function is performed in a single step via
    \begin{multline}
        \breve{\psi}(t_\text{ToF},r,z)= \mathcal{FH}^{-1}\bigg[ e^{-i\hbar \frac{q_z^2 + q_r^2}{2 m} (t_\text{ToF} - t_1)} 
            \\ \times\mathcal{FH}\left[\breve{\psi}(t_1,r,z)\right] \bigg]
        \,,
    \end{multline}
    where $\mathcal{FH}[\psi]$ descibes a Fourier transform in~$z$ direction together with a Hankel transform~\cite{Leutenegger2007Hankel} in radial direction and $q_z$ and $q_r$ are the corresponding reciprocal grids.
    \item\label{NumProcThermalExpansion_MeasureWidth} The local widths after time of flight, $\sigma(t_\text{ToF}, z)$, are then obtained by fitting a radial Gaussian at each position~$z$.
    \item Steps~\ref{NumProcThermalExpansion_AddNoise} to \ref{NumProcThermalExpansion_MeasureWidth} are typically repeated 100 times with different seed noise $\delta\rho$ and $\delta\phi$. The widths shown in Fig.~\ref{fig:_CompressionVsZ} are an average over these repetitions and additionally smoothed to account for the fact that the light sheet measurement is unable to resolve the density fluctuations seen in the simulations. The shaded areas in Fig.~\ref{fig:_CompressionVsZ} represent the standard deviation from the mean of these 100 repetitions.
\end{enumerate}

A comparison between the simulated longitudinal density $\rho_{1D}(t_\text{ToF},z)$ after ToF and the measured data is given in Fig.~\ref{fig:_Thermal_qBEC}. There we see that the simulation (blue curve) agrees well with the fitted profile for a condensate (red curve), the deviation arising from the thermal fraction missing in the simulation. The simulated density at the center appears to be too high, this could explain the discrepancy between simulation and measurement in the same region in Fig.~\ref{fig:_CompressionVsZ}.

Arguably it would be more rigorous to include the phase and density noise along with a thermal fraction right from the beginning of the simulation (\ie at trap release). In the LI-DD simulation this thermal noise would give rise to superradiance for certain parameters, mainly for dense atomic clouds (see \ref{sec:_Results} in main text). But since such a rigorous simulation would add substantial numerical cost with little additional insight for the present experiment, this endeavour is postponed for the future work.

Another reason not to include the thermal noise from the very beginning was our intention to separate the effects of superradiance and LI-DD interactions in order to show that the latter causes the atomic cloud contraction regardless of the change of the momentum distribution in the longitudinal direction.

\section{Measurement of the saturation parameter $s$}\label{sec:_measurement_of_saturation}

\begin{figure*}
\centering
\includegraphics[width=1\linewidth]{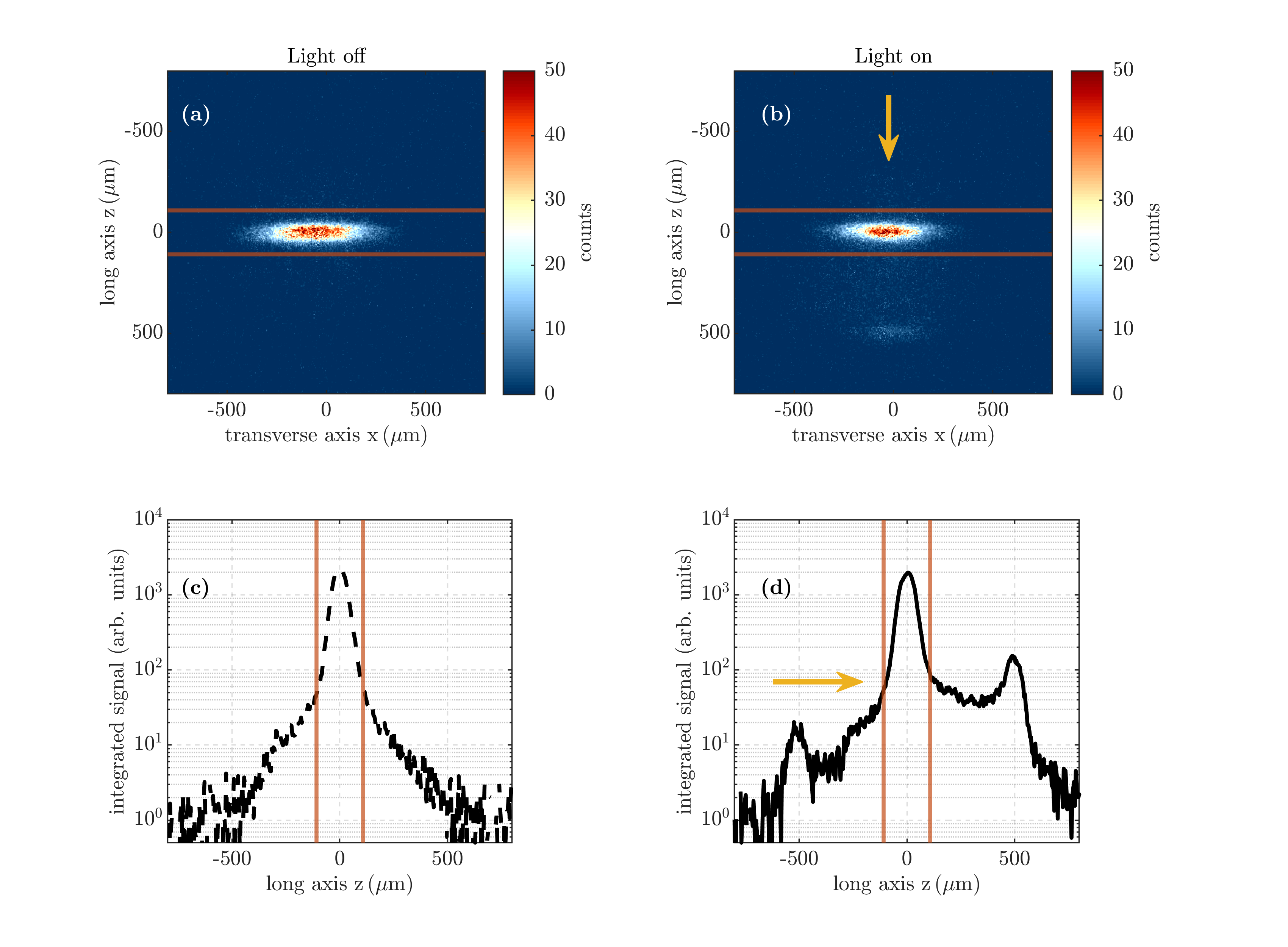}
\caption{Region of interests (ROI) for scattering rate measurement. Single shot image of a 1d-BEC after $44\,\mathrm{ms}$ time of flight without (a) and with (b) illumination with a $40\,\mathrm{\mu s}$ long laser pulse $500\,\mathrm{\mu s}$ after trap release. The data is recorded with mean atom number $N=7450(60)$, $s=319(11) \times 10^{-6}$ and $\Delta =-392 \, \Gamma$.  
The red lines indicate the ROI for the evaluation of the signal in the atom cloud (\cf Fig.~\ref{fig:_ExpansionAfterCloud_ExpAndTheory}). The laser beam is aligned parallel to the axial direction of the initial 1d-BEC and indicated by the yellow arrow. (c) \& (d):  Integrated longitudinal profile without (c) and with (d) illumination (logarithmic scale), each averaged over 5 repeats. Apart from the bulk BEC at $z=0$, two additional side-peaks at $\Delta z \sim  \pm 500\, \mathrm{\mu m}$ are visible. This is the signal from atoms that have gained a momentum  $\hbar\Delta k = \Delta z \frac{m}{t_\mathrm{ToF}} \approx \pm  2h/\lambda= \pm 2 \hbar k_L$ along $z$ while interacting with the laser beam ($\lambda\approx 780\, \mathrm{nm}$). The peaks at $\Delta z \approx \pm  500\,\mu m$, corresponding to $\Delta k \approx \pm  2 k_L$, are due to superradiant emission, mostly in the forward direction. For $0\leq k \leq 2 k_L$ ordinary single-photon scattering is clearly visible.}
\label{fig:ROI}
\end{figure*}

To measure the saturation parameter $s$ of the laser pulse, the magnetic trap is switched off and a pulse of varying duration $(0 - 200\,\mathrm{\mu s}$) is applied after $0.5\,\mathrm{ms}$ expansion time. By this time the cloud radius has expanded by a factor $\simeq \sqrt{(1+\omega_{\perp}^2t_{ToF}^2)} \approx 9.4$ and the density is low enough such that the additional losses which we have attributed to light-assisted collisions are found to be negligible in the experiment. With this setting, the total signal in the light sheet remains constant for varying pulse durations while the signal from atoms remaining in the BEC (see Fig.~\ref{fig:ROI}) decays as $S(t)=S(0) \exp{(-R \, t)}$ with single-photon scattering rate $R=\frac{\Gamma}{2}\,\frac{s}{s+1} \approx \frac{\Gamma \, s}{2}$ for small $s$. The saturation parameter $s$ of the laser pulse is obtained by an exponential fit to the signal in the bulk BEC for varying pulse duration after $500 \, \mathrm{\mu s}$ expansion time. The calibration data was taken right before or after the corresponding data sets (shown in the main text) with identical detuning and intensity.

\bibliography{LIDDI_bibliography.bib}
\end{document}